\documentclass[journal]{IEEEtran}
\usepackage{mathrsfs}
\usepackage{float}
\usepackage{xcolor}
\usepackage{textcomp}
\usepackage{amsmath}
\usepackage{cite}
\usepackage[pdftex]{graphicx}

\usepackage{tikz}
\usepackage{textcomp}
\usepackage{hyperref}
\usepackage{lipsum}

\newcommand\copyrighttext{%
  \footnotesize This work has been submitted to the IEEE for possible publication. Copyright may be transferred without notice, after which this version may no longer be accessible.}
\newcommand\copyrightnotice{%
\begin{tikzpicture}[remember picture,overlay]
\node[anchor=south,yshift=10pt] at (current page.south) {\fbox{\parbox{\dimexpr\textwidth-\fboxsep-\fboxrule\relax}{\copyrighttext}}};
\end{tikzpicture}%
}

\begin{document}

\title{Reflectarray Design Using a Discrete Dipole Framework}

\author{Aaron~V.~Diebold,
        Divya~Pande,
        Christine~Gregg,
        and~David~R.~Smith,~\IEEEmembership{Senior Member,~IEEE}
\thanks{A. Diebold, D. Pande, and D. Smith are with the Center for Metamaterials and Integrated Plasmonics, Department of Electrical and Computer Engineering, Duke University, Durham, North Carolina 27708, USA, email: aaron.diebold@duke.edu}

\thanks{C. Gregg is with the NASA Ames Research Center, Intelligent Systems Division, Moffett Field, CA 94035.}
\thanks{Manuscript received ***; revised ***.}}


\maketitle
\copyrightnotice

\begin{abstract}
We propose and numerically validate a patch reflectarray modeling approach suitable for small patches that describes each patch as a pair of polarizable magnetic dipoles. We introduce an extraction technique to obtain the effective polarizability of the patch dipoles via full-wave simulations on individual patches, followed by a beamforming design routine valid under weakly scattering configurations. This dipole framework serves as an alternative to the ray tracing model often used in reflectarray designs, in which rays are drawn from the feed point and scattered off of the patch elements. Whereas the ray tracing method solves the design problem in terms of phase delays, the dipole framework presented here has the potential to accurately design and predict beam patterns using a fully dipolar treatment of empirically characterized patches. We illustrate this technique by applying it to two modulation strategies: a variable patch size reflectarray in which the phase can be continuously tuned (\textit{grayscale} patch response), and a fixed patch size (\textit{binary} patch response) in which \textit{on}/\textit{off} modulation is achieved through selective patch electrical shorting. Methods for incorporating these cases into the dipole design framework are discussed and the results compared to those from full wave simulation.
\end{abstract}


%
\IEEEpeerreviewmaketitle

\section{Introduction}\label{sec:introduction}
\IEEEPARstart{B}{eamforming} antennas are a technological prerequisite for a variety of modern applications, with prevalent designs in the form of reflective structures such as parabolic dish antennas or collections of array elements. A reflectarray incorporates the response of both of these constituents in a planar reflection geometry that offers ease of fabrication and deployment \cite{kaddour2020foldable, beccaria2016design, hodges2017deployable}. Reflectarrays become especially appealing at higher frequencies where the use of a free space feed eliminates many of the losses and design challenges associated with microstrip feed networks \cite{huang2007reflectarray}. The beam quality achieved by reflectarray antennas as well as their compatibility with simple electronic tuning strategies \cite{kamoda201160} provide performance competitive with phased array architectures at substantially lower cost, weight, and power.

Whereas the physical realization of a beamforming reflectarray differs from that of a typical phased array, the design approach is necessarily similar. The fields for a desired beam backpropagated to the aperture plane must equate to the fields originating from the feed illumination and scattered by the reflectarray. Typically it is the phase that dominates the scattering characteristics, such that it is often a good approximation to consider only the phase distribution of the feed added to the phase distribution contributed by the reflectarray in the aperture plane. Patches situated on a ground plane are often employed to realize this phase distribution due to their well-understood electromagnetic response and relatively straightforward fabrication. 

Reflectarray design and analysis proceeds from experimental or numerical characterization of a patch over some range of tuning states \cite{pozar1997design, huang2007reflectarray}, yielding a characteristic S curve that describes the patch phase response versus geometry. The patch response can vary continuously by sweeping the patch size or, for example, using electronic modulation such as varactor diodes \cite{hum2013reconfigurable, hum2005realizing, hum2007modeling, boccia2009performance, tayebi2015dynamic}. In this case, a reflectarray can achieve \textit{grayscale} phase and amplitude tuning. Other reflectarray designs may exploit a discrete set of patch values or modulation states \cite{mei2020low, pringle2004reconfigurable, perruisseau2008monolithic, wu2008selection}, yielding quantized phase and amplitude values. At its lowest limit, a \textit{binary} reflectarray leverages only two modulation levels (\textit{on}/\textit{off}) for beamforming, accessed for instance with PIN diodes \cite{kamoda201160, carrasco2012x}.

With the phase response of the patch thus characterized, the common approach to array design proceeds by prescribing the patch distribution according to an approximate phase shift model, specifying a patch geometry at position $\mathbf{r}_i$ such that its complex response $e^{j\psi_i}$, when multiplied by the complex incident source field $e^{j\psi_{S,i}}$, equals the phase distribution required to steer a beam in the direction of $\mathbf{k}_b$, $e^{-j\mathbf{k}_b\cdot\mathbf{r}_i}$. This criterion leads to a phase distribution prescribed by \cite{pozar1997design, huang2007reflectarray, nayeri2015beam}: 
\begin{equation}
    \psi_i = -\psi_{S,i} - \mathbf{k}_b\cdot \mathbf{r}_i + 2\pi N
\label{eq:phase_mapping}
\end{equation}
where $k=2\pi/\lambda$ is the free space wavenumber, $\psi_{S,i}$ is the phase of the source field at the position of the $i$th patch, and $N$ is an integer. This design approach is simple and effective, and can be further extended by employing less restrictive assumptions. For example, while Eq. (\ref{eq:phase_mapping}) treats the feed as a point source, improved accuracy may be realized through a rigorous model of the feed radiation \cite{arrebola2009accurate}. The work in \cite{baladi2021dual} follows this approach for the case of a circularly polarized, reconfigurable reflectarray. \cite{florencio2015reflectarray} illustrates how introducing more degrees of freedom enables broadband designs with sufficient polarization control to meet stringent telecom requirements. In addition, that work illustrates that as the patch geometry gets increasingly more complex, the design procedure of Eq. (\ref{eq:phase_mapping}) must be supplemented or replaced by more sophisticated computational approaches such as finite element methods (FEM) or the method of moments (MoM) \cite{carver1981microstrip}.

While FEM and MoM approaches can be extremely flexible and accurate, the computational requirements can become excessive for large problems with widely disparate feature sizes, as is often the case for reflectarrays with small, application-tailored unit cells \cite{pozar2007wideband}. Motivated by the standard cavity model of a microstrip patch antenna as a pair of polarizable magnetic dipoles, we instead seek to take an effective medium approach and describe the collective array response through a discrete dipole formalism. The discrete dipole approximation (DDA) is a technique for modelling scattering by small ($<\lambda/2$) ``particles.'' It offers computational savings through the use of an analytically defined Green's function \cite{yurkin2007discrete} and a consequent reduction in the number of electromagnetic degrees of freedom, as well as compatibility with fast Fourier transform acceleration strategies \cite{goodman1991application}. This leads to drastic reduction in computation time for large-scale problems---full-wave simulations that take hours to compute can instead be completed in seconds using the DDA. Originally formulated for polarizable electric dipoles \cite{purcell1973scattering, draine1994discrete}, the DDA is now a mature framework for describing fully coupled electric and magnetic dipole systems \cite{mulholland1994light, bowen2012using}.

In the context of antenna design, the polarizability design framework has been widely developed and applied to metamaterials and metasurfaces consisting of subwavelength elements \cite{imani2020review}. For electrically small elements, the scattering is predominantly dipolar in nature, and higher order contributions may be neglected. These same arguments apply to the cavity model of a subwavelength patch on a thin dielectric substrate. If the scattering response of the metamaterial or patch element is sufficiently dipolar, one can avoid the computational complexity of finite element or moment methods that require computing field solutions over the entire aperture and surrounding space. Instead, one can rigorously predict scattering behavior from arbitrary collections of dipolar elements using the DDA. Modelling a collection of dipoles in this way requires only the scattering response (polarizability) of each element in addition to the domain propagation behavior described in terms of a Green's function. The forward problem of computing the fields for a given set of polarizabilities has been validated for waveguide slots in \cite{pulido2016discrete, pulido2017polarizability} and for more general metasurfaces and metamaterial structures in \cite{pulido2017polarizability, bowen2012using, landy2014two, yoo2019analytic, yoo2019analytic2, f2016analytical}. 

The inverse problem of selecting the polarizabilities that achieve a desired output pattern is generally nonlinear due to mutual coupling effects. One approach to achieving a design solution is to incorporate coupling using the coupled dipole model and proceed iteratively to arrive at a satisfactory set of polarizabilities as in \cite{yoo2019analytic, pande2020symphotic}. Otherwise, if mutual coupling can be minimized through, for example, sufficient separation between the array elements, one can apply a Born approximation and neglect mutual coupling effects in the design procedure to arrive at a so-called ``holographic'' solution. The analysis of a beamforming metasurface in \cite{smith2017analysis} follows this approximation. Furthermore, it is reported in \cite{pozar1993analysis} that mutual coupling indeed has little observed effect in the authors' practical reflectarray design example. For these reasons, and with the goal of first reconciling the reflectarray design process with the DDA, we will seek a holographic expression for the required polarizabilities in a patch reflectarray by neglecting mutual coupling between patches and operating in the weakly coupled, dipolar element regime ($>\lambda/4$ spacing, $<\lambda/2$ element size). Rigorous accommodation of electromagnetic dipolar coupling in a reflectarray will be the topic of future studies.
 
We will outline the polarizability description of patches in the following section, followed by a polarizability retrieval strategy for patch characterization in Section \ref{sec:polarizability_retrieval}. Section \ref{sec:holographi_design} describes a holographic reflectarray design procedure formulated within the discrete dipole model. Finally, we will provide numerical results illustrating the implementation of this procedure and the effect of illumination diversity on beam performance by applying the discrete dipole methodology to grayscale and binary patch reflectarrays.

\section{Polarizable Patch Model}\label{sec:patch_model}

\begin{figure}[ht]
\centering
\includegraphics[width=.8\columnwidth]{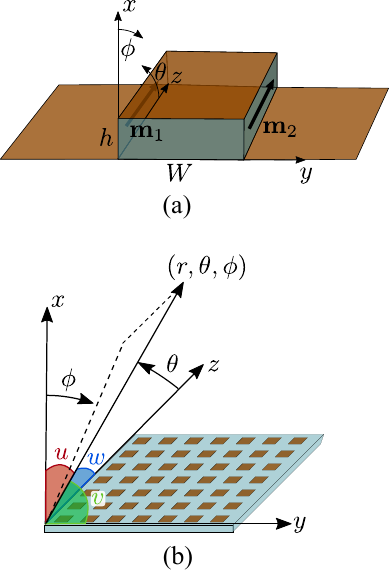}
\caption{Patch reflectarray geometry and coordinate system used in this analysis. (a) A single ground plane-backed patch and the associated spherical coordinate system used to describe its radiated fields. (b) A patch reflectarray oriented in the $yz$ plane. $u$, $v$, and $w$ are direction cosines associated with the $x$, $y$, and $z$ axes, respectively.}
\label{fig:coordinates}
\end{figure}

A standard description of a ground plane-backed microstrip patch antenna treats the gap beneath the patch as a cavity with perfect electric conductors (PEC) as walls on the patch and ground plane surfaces, and perfect magnetic conductors (PMC) as the vertical walls over the patch ``slots''. These boundary conditions give rise to a collection of cavity modes that may be excited inside the gap between the patch and ground plane. An incident electromagnetic field illuminating the patch can couple to these cavity modes to excite fields in the gap that in turn can be described by radiating effective magnetic surface currents over each of the four patch slot surfaces. If we assume that an incident magnetic field is directed along the $z$ axis and excites only the lowest order cavity mode, then the fields from the faces oriented along the $xy$ plane (see Figure \ref{fig:coordinates}(a)) cancel, leaving two radiating slots described by magnetic surface currents $\mathbf{J}_{m1}$ and $\mathbf{J}_{m2}$. These magnetic surface currents can be defined according to surface equivalence principles by \cite{balanis2015antenna}
\begin{equation}
    \mathbf{J}_{mk}= \mathbf{E}_k\times\hat{n}
\label{eq:surface_equivalence}
\end{equation}
where $\mathbf{E}_k$ is the total electric field over the surface of slot $k$. For sufficiently small patch dimensions relative to the wavelength, these magnetic currents can be approximated as point magnetic dipoles $\mathbf{m}_k$ for $k=1,2$ that are related to the continuous magnetic surface currents by 
\begin{equation}
    \mathbf{J}_{mk}(\mathbf{r})=j\omega \mu_0 \mathbf{m}_k \delta(\mathbf{r}-\mathbf{r}_k).
\label{eq:point_dipole}
\end{equation}
at angular frequency $\omega$, with $\mu_0$ the magnetic permeability of free space.

As described in \cite{bowen2016theory}, these radiating slots form polarizable magnetic dipoles, and can be equivalently characterized by a magnetic polarizability tensor $\bar{\bar{\alpha}}_{k}$ in which each nonzero component follows a Lorentizan resonance profile with resonance frequency $\omega_0$:
\begin{equation}
\left[\bar{\bar{\alpha}}_{k}\right]_{mn}=\frac{F\omega^2}{\omega_0^2-\omega^2+j\gamma\omega}
\label{eq:lorentzian}
\end{equation}
with oscillator strength $F$ and damping factor $\gamma$ related to the oscillator quality factor $Q$ by $\gamma=\omega_0/Q$. The polarizability tensor $\bar{\bar{\alpha}}_{k}$ is in turn related to the magnetic dipole moment through $\mathbf{m}_k=\bar{\bar{\alpha}}_{k} \mathbf{H}_t$, where we have chosen to define the induced magnetic dipole moment describing the patch slot response as proportional to the total illuminating magnetic field $\mathbf{H}_t$, given by the superposition of the illuminating magnetic field $\mathbf{H}_i$ with its reflection: $\mathbf{H}_t = (1-\Gamma)\mathbf{H}_i$. Here, we take $\Gamma$ to be the appropriate reflection coefficient for the dielectric-ground plane substrate and magnetic field polarization under a plane wave approximation, though an improved estimate may be obtained through more rigorous numerical methods. The slot's magnetic polarizability can be considered the fundamental quantity characterizing the electromagnetic response of a dipolar patch with a given geometry, in a given electromagnetic environment. We will assume that the patches are geometrically symmetric in such a way that the polarizability of the two radiating slots, $\bar{\bar{\alpha}}_1$ and $\bar{\bar{\alpha}}_2$, are equal.

Using the polarizability model, the fields $\mathbf{H}$ over an array of $N$ dipoles arising from an incident magnetic field $\mathbf{H}_t$ obey a set of coupled linear equations \cite{bowen2012using, pulido2017polarizability}:
\begin{equation}
    \mathbf{H}(\mathbf{r}_p)=\mathbf{H}_t(\mathbf{r}_p) + \sum_{q\neq p} \bar{\bar{G}}(\mathbf{r}_p, \mathbf{r}_q)\bar{\bar{\alpha}}(\mathbf{r}_q)\mathbf{H}(\mathbf{r}_q)
\label{eq:DDA}
\end{equation}
where $p, q = 1,...,N$ and $\bar{\bar{G}}(\mathbf{r}_p, \mathbf{r}_q)$ is the free-space dyadic Green's function \cite{collin1990field}. These equations can be simultaneously solved to recover the total fields $\mathbf{H}(\mathbf{r}_p)$, including mutual coupling, at the discrete element locations $\mathbf{r}_p$. To reduce complexity and achieve a holographic solution, we will assume weak scattering in the following, in which case $\mathbf{H}(\mathbf{r}_p) = \mathbf{H}_t(\mathbf{r}_p)$.

\section{Patch Polarizability Retrieval}
\label{sec:polarizability_retrieval}
Using the description of a patch as a set of two radiating dipoles, the response of a patch reflectarray can be decomposed into the superposition of (1) the radiation from the collection of magnetic dipoles over the antenna with (2) the fields reflected from the ground plane. While modelling ground plane reflections is straightforward, determined by the illumination profile and the ground plane material properties, accurate modelling of the radiation from each patch dipole requires a method for characterizing the slot polarizabilities. Related formulations describing patches and metamaterial elements offer analytic descriptions using, for example, notions of impedance and equivalent circuit representations \cite{carver1981microstrip, lee1985equivalent, hum2007modeling, luukkonen2008simple, costa2012closed, tayebi2015dynamic, hum2017equivalent}, though these typically yield the element phase response in terms of a reflection coefficient that is not readily compatible with the DDA model. Analytic expressions for the polarizability of metamaterial nanopatches are given in \cite{bowen2016theory} and can potentially be adapted to radio frequency (RF)-scale microstrip patches. Here, in the interest of flexible and robust performance, we develop numerical and experimental characterization strategies based on measurement inversion. \cite{pulido2017polarizability, scher2009extracting}. Developed here for linearly polarized rectangular patches, similar techniques can be employed to extend the formulations to more complex elements represented by arbitrary polarizability tensors \cite{liu2016polarizability}. Once the patch polarizability is retrieved, patch far field contributions can be predicted through standard radiation integrals. In the following, we introduce and compare two methods for characterizing patch polarizabilities.

\subsection{Polarizability Extraction by Surface Equivalence}
The first method for quantitative polarizability extraction assumes that the magnetic surface currents $\mathbf{J}_{mk}$ can be calculated over the slot surfaces through, for instance, direct evaluation of Eq. (\ref{eq:surface_equivalence}) in numerical simulations. Then the magnetic dipole moment of the point magnetic dipole describing a single slot can be obtained upon integrating Eq. (\ref{eq:point_dipole}) over the slot surface as
\begin{equation}
    \mathbf{m}_k = \frac{1}{j\omega \mu_0}\int \mathbf{J}_{mk}(\mathbf{r})d^2\mathbf{r}.
\end{equation}
Consistent with the geometry of Fig. \ref{fig:coordinates}(a) in which the slot dipoles are oriented in the $z$ direction, we can then recover the $z$ component of the polarizability corresponding to the $k$th slot as
\begin{equation}
    \alpha_{kz} = \frac{m_{kz}}{H_{tz}(\mathbf{r}_0)}
\label{eq:slot_extraction}
\end{equation}
where $H_{tz}(\mathbf{r}_0)$ is the total (incident plus reflected) magnetic field evaluated at the center of the slot. Assuming patch symmetry, $\alpha_{1z}=\alpha_{2z}$, and only one slot need be evaluated in this fashion.

\subsection{Far-Field Polarizability Extraction}
A second approach for polarizability extraction recovers the patch polarizability from far field electric field measurements and thus is amenable to experimental remote measurement techniques. The far field electric field radiated by a magnetic dipole oriented in the $z$ direction is given by:
\begin{equation}
E_{\phi,1} = \frac{-\omega k \mu_0}{4\pi r} \alpha_z H_{tz}(\mathbf{r}_0)e^{-jkr}\textrm{sin}\theta.
\end{equation}
Applying this to the two radiating dipoles comprising a single patch yields the total electric field radiated by the patch as
\begin{equation}
\begin{aligned}
\label{eq:patch_far_field}
    E_{\phi} &= \frac{-\omega k \mu_0}{4\pi r} e^{-jkr}\alpha_z (1-\Gamma) \textrm{sin}\theta\; 2\textrm{cos}\left(\frac{kh}{2}\textrm{sin}\theta \textrm{cos}\phi\right)\\&\times\left[ H_{iz}(\mathbf{r}_1) e^{-\frac{jkW}{2}\textrm{sin}\theta\textrm{sin}\phi} + H_{iz}(\mathbf{r}_2) e^{\frac{jkW}{2}\textrm{sin}\theta\textrm{sin}\phi}\right],
\end{aligned}
\end{equation}
where the factor $2\textrm{cos}\left(\frac{kh}{2}\textrm{sin}\theta \textrm{cos}\phi\right)$ is an array factor term resulting from the superposition of a single slot dipole with its image in the ground (PEC) plane. As defined previously, $\Gamma$ is the reflection coefficient from the dielectric-ground plane substrate, which can be computed locally under a plane wave approximation or via more rigorous numerical methods, and $W$ is the patch width separating the radiating dipoles. If the incident illumination is arranged so as to yield equal magnetic field values at the two slots ($H_{iz}(\mathbf{r}_1) = H_{iz}(\mathbf{r}_2)=H_{iz}$), e.g. for a normally incident plane wave, and if the substrate thickness $h$ is small, then Eq. \ref{eq:patch_far_field} reduces to
\begin{equation}
    E_{\phi} = \frac{-\omega k \mu_0}{\pi r} e^{-jkr} \alpha_z (1-\Gamma) H_{iz}\textrm{sin}\theta \textrm{cos}\left(\frac{kW}{2} \textrm{sin}\theta \textrm{sin}\phi\right)
\label{eq:ff_extraction}
\end{equation}
where now the term $2\textrm{cos}\left(\frac{kW}{2} \textrm{sin}\theta \textrm{sin}\phi\right)$ gives the array factor for the two patch dipoles. Given a complex scattered electric field measurement $E_{\phi}$, this expression can be inverted to obtain the slot polarizability $\alpha_z$. As this expression is comprised of the electric field radiated only by the patch dipoles, direct reflection from the ground plane into the far field is not included (only its interaction with the patch dipoles is contained in Eq. (\ref{eq:ff_extraction})). Therefore, numerical or experimental implementation requires that the patch response be first isolated through a subtraction of the ground plane reflection.

\begin{figure}[ht]
\centering
\includegraphics[width=\columnwidth]{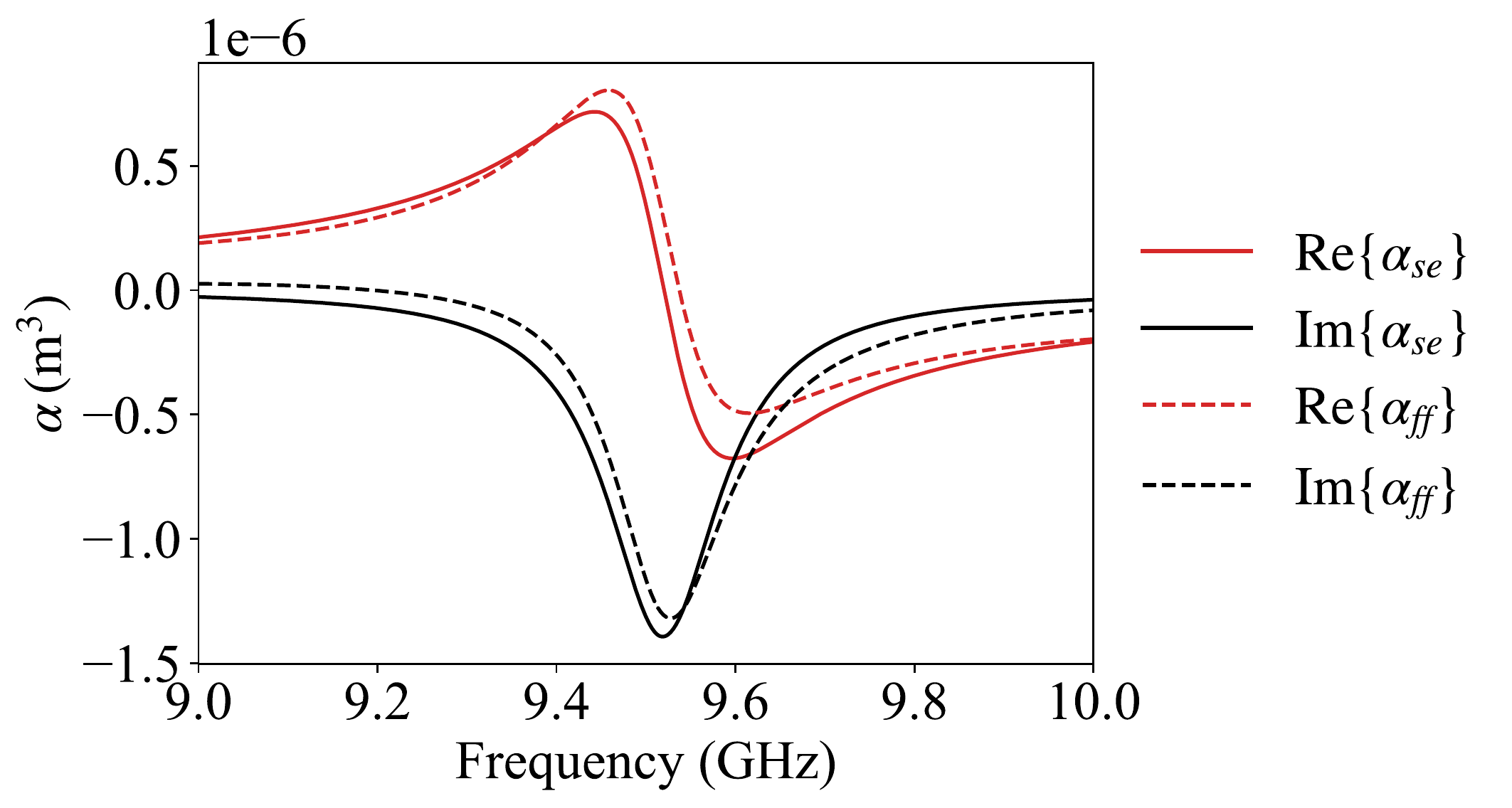}
\caption{Comparison of polarizability values extracted from simulated data using the proposed surface equivalence extraction method ($\alpha_{se}$) and far-field inversion ($\alpha_{ff}$). The patch under study is 1.5 cm $\times$ 1.5 cm and separated from the ground plane by 0.25 mm of air.}
\label{fig:extraction_comparison}
\end{figure}

We compare these two polarizability extraction methods applied to a square patch of size 1.5 cm $\times$ 1.5 cm in Fig. \ref{fig:extraction_comparison}. In this example, the patch is separated from the ground plane by 0.25 mm of air, and the polarizability values obtained using the two methods show close agreement. The small discrepancies may arise from meshing errors in the distinct simulation methods.

\begin{figure}[ht]
\centering
\includegraphics[width=\columnwidth]{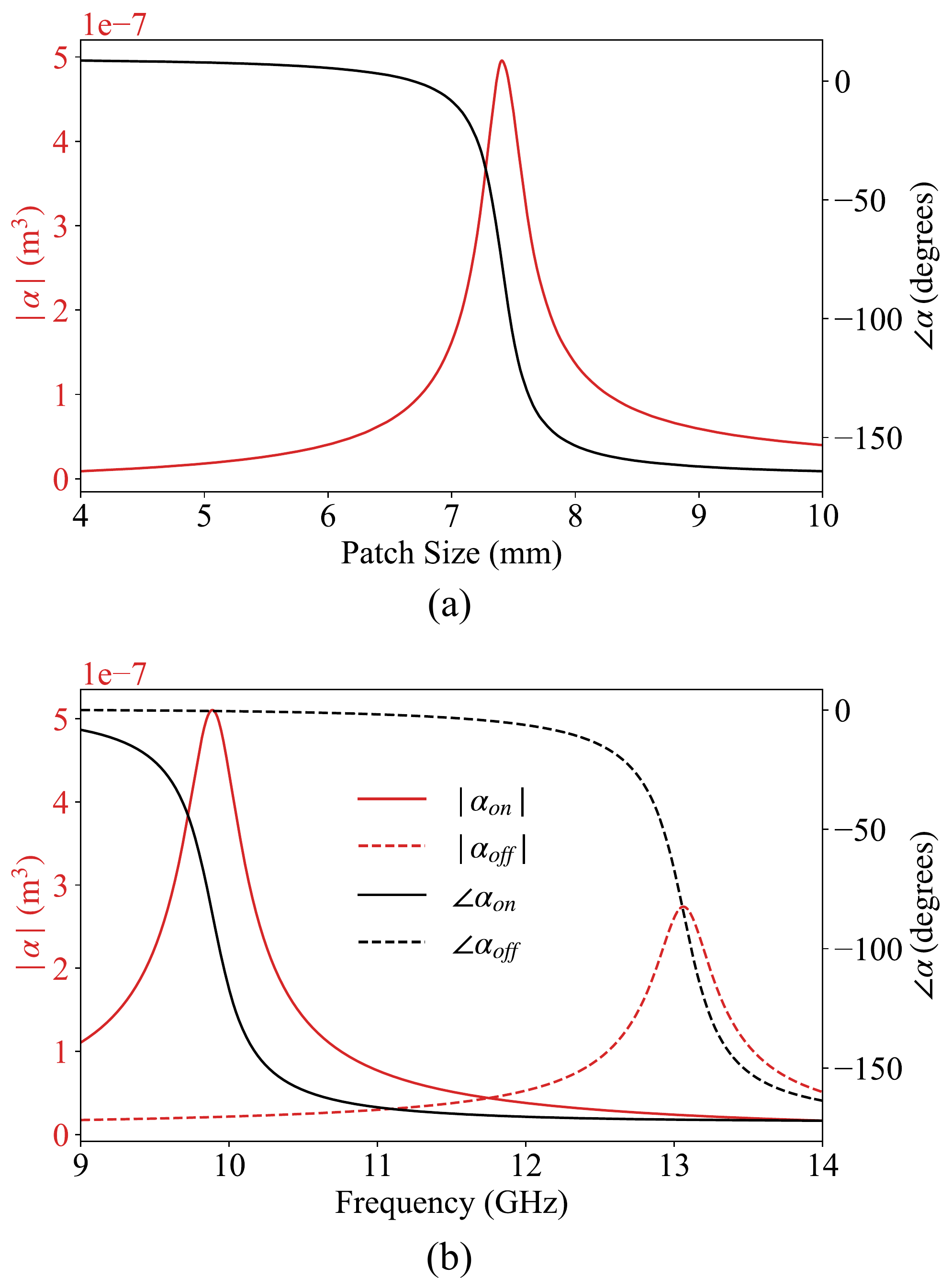}
\caption{(a) Magnitude and phase of magnetic polarizability values for square patches on 0.762 mm-thick Rogers 4350B dielectric, ranging in length from 4 mm to 10 mm. The polarizability values were retrieved from numerical integration of equivalent surface currents according to Eq. (\ref{eq:slot_extraction}). (b) Magnetic polarizability values for \textit{on} (radiating) and \textit{off} (non-radiating) patch states. The patch size is 7.5 mm $\times$ 7.5 mm.}
\label{fig:polarizability}
\end{figure}

Simulated results obtained using the surface equivalence polarizability retrieval method are given in Fig. \ref{fig:polarizability} for square patches on 0.762 mm-thick Rogers 4350B substrate ($\epsilon_r=3.48,\: \textrm{tan}\delta=0.0037$). Figure \ref{fig:polarizability}(a) illustrates the resulting polarizability for patches of varying size at a frequency of 10 GHz. As the patches range in length from 4 mm to 10 mm, the polarizability sweeps out a resonance that peaks at 7.4 mm. As demonstrated in many patch reflectarray examples, and as we will similarly demonstrate under our polarizability framework in Section \ref{sec:results}, such a library of patch geometries can be used to design a grayscale reflectarray by properly selecting the patch geometry at each array position in order to conform to some design criteria. 

The polarizability values for a 7.5 mm-wide ($\lambda/4$ at 10 GHz) square patch are plotted in Fig. \ref{fig:polarizability}(b) as a function of frequency. Here, in contrast to the grayscale values accessed in Fig. \ref{fig:polarizability}(a), we explore a simple method for achieving two binary patch states. The radiating state corresponds to the 7.5 mm-wide square patch with a polarizability that peaks just below 10 GHz. The patch can be made approximately non-radiating at 10 GHz by shorting it to ground, thus shifting its resonance far from the operating frequency. In this case, we connect a metallized via from the patch to the ground plane at the center of each radiating slot (see Fig. \ref{fig:PM-CST}(a)). We refer to the radiating and non-radiating states of this binary design as $\alpha_{\textrm{on}}$ and $\alpha_{\textrm{off}}$, respectively. We will investigate the feasibility of utilizing such a binary scheme in later studies. In practice, dynamic control of patch response can be achieved through mechanical or electronic strategies \cite{nayeri2015beam}, for example by incorporating electronic tuning elements into a fixed patch geometry. These approaches can use PIN diodes for binary tuning or varactor diodes for grayscale tuning.

\section{Holographic Array Design}\label{sec:holographi_design}
The above procedures illustrate methods for extracting or characterizing the magnetic polarizabilities that define a patch with a given geometry. Through the characterization of patch responses over a range of geometries or tuning states, a complete design set can be obtained. Assuming such a set of values is known, an inverse design problem can be solved: given a desired output far-field beam, determine the required polarizabilities over an aperture. This is the analogue of the objective typically addressed through the simple phase mapping of Eq. (\ref{eq:phase_mapping}) in conventional reflectarray design. Here, instead of describing the coupling of the incident fields to patches in terms of an abstract phase response, we use the polarizability description to arrive at an electromagnetic field theoretical solution. 

\begin{figure}[ht]
\centering
\includegraphics[width=\columnwidth]{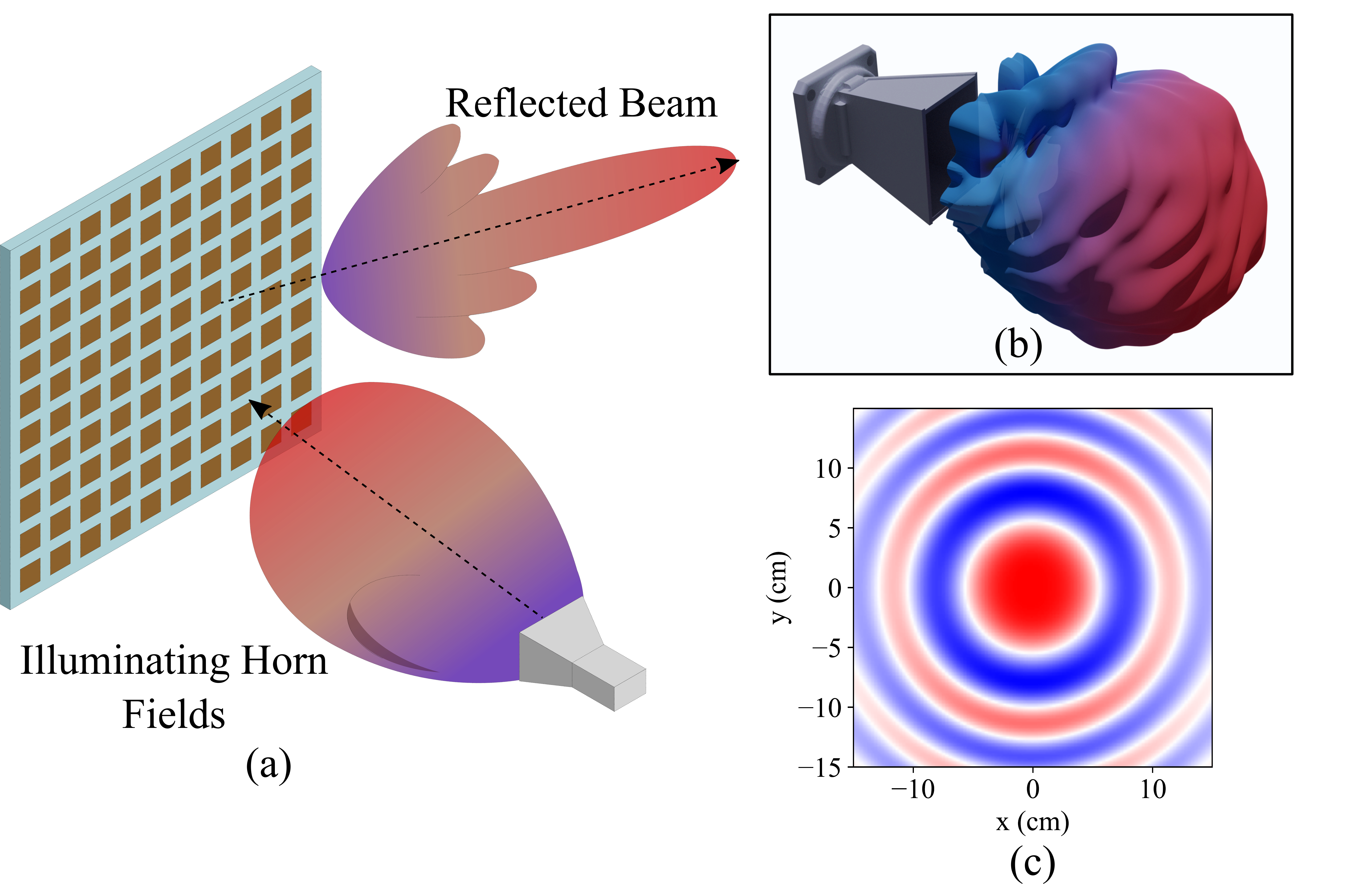}
\caption{(a) Illustration of patch reflectarray antenna employing horn illumination. (b) A rendering of the pyramidal horn antenna used for the studies in this work, with an example of its radiation pattern superimposed. (c) Real part of the magnetic field from the pyramidal horn in (b) computed at 10 GHz over a 30 cm $\times$ 30 cm reflectarray surface. The distance from the horn to the reflectarray surface is 20 cm.}
\label{fig:schematic}
\end{figure}

To design a beamforming reflectarray antenna, we seek an array of patches described by a spatial polarizability distribution which, when combined with ground plane reflections, transforms an incident illumination profile to a beam in a specified direction (see Fig. \ref{fig:schematic}(a)). To achieve this result, the realized polarizability values will necessarily depend on the form of illumination, though the design approach can generally be applied for any illumination profile. A commonly used illumination source that we adopt for the results that follow is a horn antenna. Figure \ref{fig:schematic}(b) shows a rendering of a 10 dBi X-band pyramidal horn antenna (PE9856B-10) along with a plot of its radiation pattern measured at 10 GHz. Using a standard analytical description of this pyramidal horn \cite{balanis2015antenna}, the fields illuminating the reflectarray ($\mathbf{H}_i$) can be calculated numerically, as shown in Fig. \ref{fig:schematic}(c) at 10 GHz and with the horn placed 20 cm from the reflectarray surface. Combined with an expression for the output beam, one can then prescribe the reflectarray design in terms of the required polarizabilities at each array position.

We start by expressing the far fields of our reflectarray antenna in terms of the electric and magnetic surface currents. The far-field magnetic field radiated by our reflectarray antenna can be decomposed into contributions from the magnetic and electric vector potentials $\mathbf{A}$ and $\mathbf{F}$, respectively \cite{balanis2012advanced}:
\begin{equation}
\begin{aligned}
\label{eq:H_ff1}
    \mathbf{H} &= \mathbf{H}_A + \mathbf{H}_F \\
    &= \frac{-j\omega}{\eta}\hat{k}_b \times \mathbf{A} - j\omega \mathbf{F}
\end{aligned}
\end{equation}
where $\hat{k}_b$ is the unit vector in the beam direction and $\eta=\sqrt{\frac{\mu_0}{\epsilon_0}}$ is the free space impedance. The magnetic and electric vector potentials are related, respectively, to the Fourier transforms of the electric and magnetic surface currents over the antenna surface:
\begin{gather}
    \mathbf{A}=\frac{\mu_0 e^{-jkr}}{4\pi r} \int\int\mathbf{J}_e e^{j\mathbf{k}\cdot\mathbf{r}'}d^2\mathbf{r}' = \frac{\mu_0 e^{-jkr}}{4\pi r} \mathscr{F}\{\mathbf{J}_e\} \\ 
    \mathbf{F}=\frac{\epsilon_0 e^{-jkr}}{4\pi r} \int\int\mathbf{J}_m e^{j\mathbf{k}\cdot\mathbf{r}'}d^2\mathbf{r}' = \frac{\epsilon_0  e^{-jkr}}{4\pi r} \mathscr{F}\{\mathbf{J}_m\}
\end{gather}
where $\mathscr{F}\{\cdot\}$ denotes the two-dimensional Fourier transform over the aperture coordinates. Note that we are neglecting finite aperture effects for the sake of deriving a design expression. In reality, a finite aperture implies a varying electromagnetic environment dependent on an element's position within the array due to mutual coupling effects \cite{bhattacharyya2006phased}. As we are operating under a weakly coupled approximation, the dominant effect omitted by this approximation will be beam broadening, which will arise naturally upon practical, numerical implementation.  Inserting these definitions into Eq. (\ref{eq:H_ff1}) gives an expression relating the far-field magnetic field to the electric and magnetic surface currents over the antenna surface:
\begin{equation}
\begin{aligned}
\label{eq:H_ff2}
    \mathbf{H} &= \frac{-j\omega\mu_0 e^{-jkr}}{\eta 4\pi r} \mathscr{F}\{\hat{k}_b\times\mathbf{J}_e\} - \frac{j\omega \epsilon_0 e^{-jkr}}{4\pi r} \mathscr{F}\{\mathbf{J}_m\} \\
    &= \frac{-j\omega \epsilon_0 e^{-jkr}}{4\pi r} \left[ \eta \mathscr{F}\{\hat{k}_b \times \mathbf{J}_e\} + \mathscr{F}\{\mathbf{J}_m\}\right].
\end{aligned}
\end{equation}
An ideal far-field beam under the infinite aperture approximation should satisfy
\begin{equation}
    \mathbf{H} = \mathbf{H}_0 \delta(\mathbf{k} - \mathbf{k}_b).
\end{equation}
That is, the magnetic field profile in the far field of the proposed infinite array should peak in the desired beam direction $\mathbf{k}_b$. Equating this desired response to the expression (\ref{eq:H_ff2}), taking an inverse Fourier transform, and using the Fourier transform relationship between a delta function and a plane wave yields a prescription for the magnetic and electric surface currents required to form a beam in the far field:
\begin{equation}
\label{eq:holographic_design}
    \mathbf{J}_m + \eta \hat{k}_b\times \mathbf{J}_e = a\mathbf{H}_0 e^{-j\mathbf{k}_b\cdot \mathbf{r}'}
\end{equation}
where $a=\frac{j\kappa 4\pi r}{\omega\epsilon_0 e^{-jkr}}$ is a constant incorporating the far field distance $r$ and normalization constant $\kappa$, and provides some scaling freedom in the holographic design procedure.

The magnetic surface currents $\mathbf{J}_m$ consist of the discrete patch dipoles:
\begin{equation}
\label{eq:Jm}
    \mathbf{J}_m = \frac{j\omega \mu_0}{\Lambda^2}\bar{\bar{\alpha}}\mathbf{H}_t
\end{equation}
with patch spacing $\Lambda$. The illuminating fields also excite electric surface currents $\mathbf{J}_e$ over the substrate that can be defined through surface equivalence principles as
\begin{equation}
\label{eq:Je}
    \mathbf{J}_e = \hat{n} \times \left[(1-\Gamma)\mathbf{H}_i\right] = \hat{n} \times \mathbf{H}_t
\end{equation}
where $\hat{n}$ denotes the antenna surface normal. Finally, using these definitions for the electric and magnetic surface currents, one can show for a diagonal polarizability tensor that the $j$th component along the diagonal is given by
\begin{equation}
\begin{aligned}
    \alpha_j = \frac{-j\Lambda^2}{\omega \mu_0} &\frac{1}{\mathbf{H}_t \cdot \hat{j}} \bigg[a(\mathbf{H}_0 \cdot \hat{j})e^{-j\mathbf{k}_b \cdot \mathbf{r}'} \\ &- \eta \left((\hat{k}_b\cdot\mathbf{H}_t)(\hat{j}\cdot\hat{n}) - (\hat{k}_b\cdot\hat{n})(\mathbf{H}_t\cdot\hat{j})\right)\bigg].
\end{aligned}
\end{equation}
For the geometry shown in Fig. \ref{fig:coordinates} with the magnetic field oriented in the $z$ direction and the antenna in the $yz$ plane, the required $z$ component of the polarizability is thus given by
\begin{equation}
\label{eq:polarizability_design}
    \alpha_z = \frac{-j\Lambda^2}{\omega \mu_0} \left[\frac{a H_{0z}}{H_{tz}}e^{-j\mathbf{k}_b\cdot\mathbf{r}'} + \eta (\hat{k}_b\cdot\hat{n})\right].
\end{equation}

\subsection{Lorentzian Polarizability and the Patch Phase Response}
Equation (\ref{eq:holographic_design}) provides the physical foundation for beamforming with a reflectarray by decomposing a plane wave over the aperture into the superposition of electric currents arising from ground plane reflections and magnetic currents describing the radiating patches. The required patch geometries can then be prescribed by the corresponding polarizabilities given in Eq. (\ref{eq:polarizability_design}). In view of the Lorentzian form of the patch polarizabilities (Eq. (\ref{eq:lorentzian})), one notices that the available patch slot contributions traverse only 180 degrees of phase (Fig. \ref{fig:polarizability}), in contrast to the full 360 degrees of phase shift accessible to reflection-mode patches presented in the literature \cite{huang2007reflectarray}. Here, we examine Eq. (\ref{eq:holographic_design}) in more detail in order to reconcile this different behavior. 

In order to simplify the analysis, we suppose that the antenna is aligned with the $yz$ plane and illuminated by a magnetic field linearly polarized in the $z$ direction, so that $\mathbf{H}_t=H_t(\mathbf{r}')\hat{z}$ and $\hat{n}=\hat{x}$. In addition, we assume that the radiating patch slots are similarly oriented in the $z$ direction, resulting in a single nonzero element of the polarizability tensor $\alpha_z$. This geometry leads to electric and magnetic surface currents given by
\begin{subequations}
\begin{align}
&\mathbf{J}_e=-H_t\hat{y}, \label{eq:Je_reduced}\\
&\mathbf{J}_m=\frac{j\omega\mu_0}{\Lambda^2}\alpha_z H_t\hat{z}. \label{eq:Jm_reduced}
\end{align}
\end{subequations}
Further restricting our present analysis to a broadside beam with $\mathbf{H}_0=H_0\hat{z}$, we have $\hat{k}_b=\hat{x}$, and the electric surface current term on the left hand side of Eq. (\ref{eq:holographic_design}) is given by $\eta\hat{k}_b\times\mathbf{J}_e=-\eta H_t\hat{z}$.

\begin{figure}[ht!]
\centering
\includegraphics[width=.7\columnwidth]{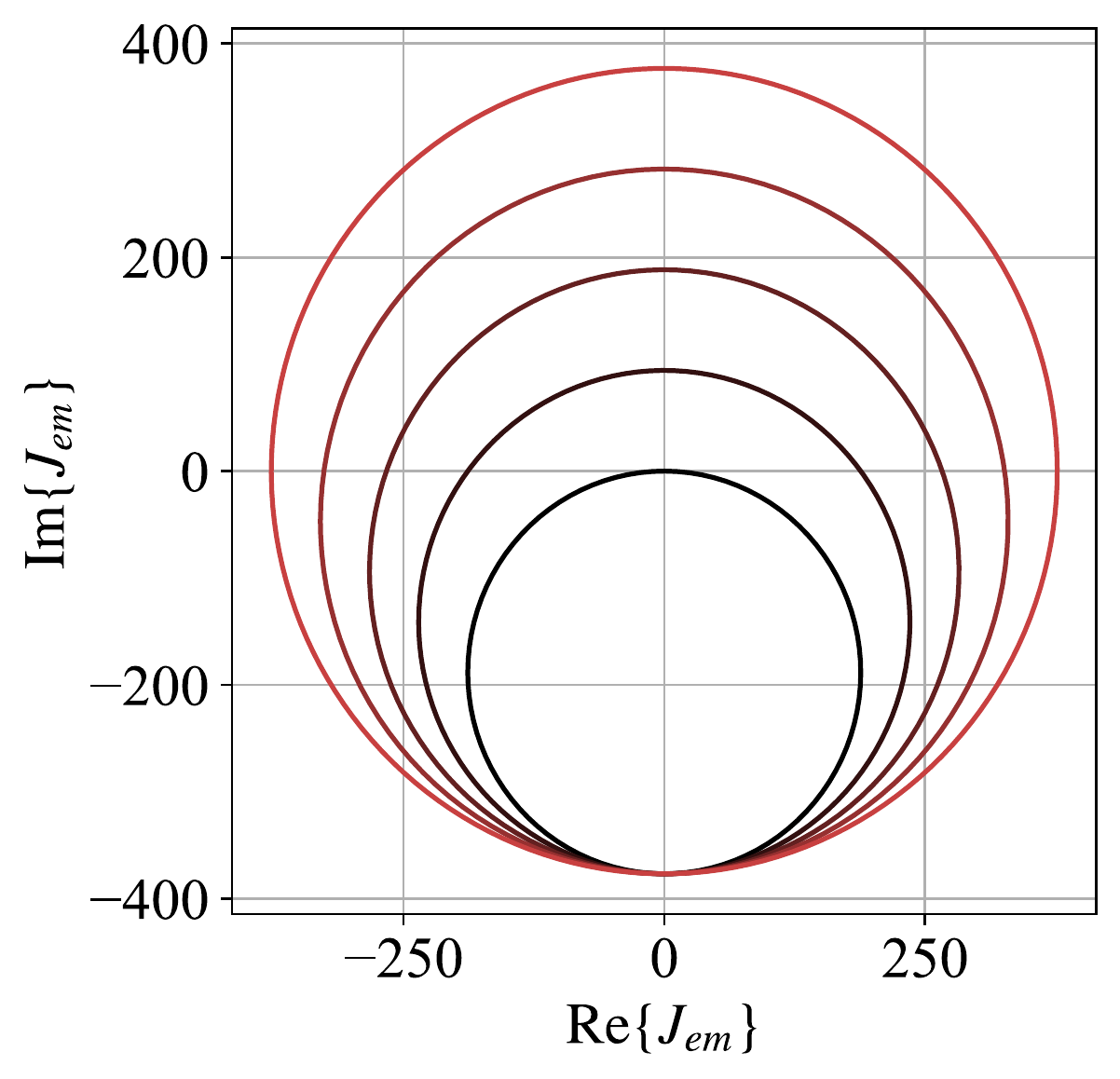}
\caption{Complex polar plots of ($j$ times) the left hand side of Eq. (\ref{eq:holographic_design_reduced}), denoted $J_{\textrm{em}}$, characterizing the combined electromagnetic response of a patch, as the magnitude of the patch polarizability varies from $\alpha_{0,\textrm{opt}}/2$ (black) to $\alpha_{0,\textrm{opt}}$ (red), where $\alpha_{0,\textrm{opt}}=\frac{2\Lambda^2}{k_0}$ is the polarizability magnitude resulting in optimal phase coverage. The expression was multiplied by $j$ in order to highlight the connection with the constrained Lorentzian polarizability profile.}
\label{fig:circle_shift}
\end{figure}

We will now confine the study to a single frequency and parameterize the slot polarizabilities in a more convenient form:
\begin{equation}
    \label{eq:polarizability_in_complex_plane}
    \alpha_z = \alpha_0\left(\frac{-j + e^{-j\psi}}{2}\right),
\end{equation}
where $\alpha_0$ is a constant magnitude and $\psi$ a phase ranging from 0 to 360 degrees. In the complex plane, Eq. (\ref{eq:polarizability_in_complex_plane}) represents the constrained form of a Lorentzian resonator as a circle shifted downward by its radius, so that its actual phase response only covers 180 degrees. Using Eq. (\ref{eq:polarizability_in_complex_plane}) in Eq. (\ref{eq:Jm_reduced}) gives
\begin{align}
    \mathbf{J}_m &= \frac{j\omega \mu_0} {\Lambda^2}\left(\frac{-j+e^{-j\psi}}{2}\right)H_t\hat{z} \\
    &=\frac{\alpha_0\eta k_0}{2\Lambda^2}\left(1+je^{-j\psi}\right)H_t\hat{z},
\end{align}
and the holographic design equation (\ref{eq:holographic_design}) becomes
\begin{equation}
    \left[\left(b-\eta\right) + jb e^{-j\psi}\right]H_t=aH_0e^{-j\mathbf{k}_b\cdot\mathbf{r}'}
\label{eq:holographic_design_reduced}
\end{equation}
where $b=\frac{\eta\alpha_0 k_0}{2\Lambda^2}$ and $a$ is defined above. Examining Eq. (\ref{eq:holographic_design_reduced}), we see that the electric currents exactly compensate for the offset of the Lorentzian circle in the complex plane when $b=\eta$, or when 
\begin{equation}
\alpha_{0,\textrm{opt}}=\frac{2\Lambda^2}{k_0}.
\label{eq:optimal_polarizability}
\end{equation}
In this case, the combined electromagnetic patch response represented by the left hand side of Eq. (\ref{eq:holographic_design_reduced}) becomes proportional to $e^{-j\psi}$ and recovers a full 360 degrees of phase shift. Figure \ref{fig:circle_shift} illustrates the convergence to an ideal phase profile realized by the patch electromagnetic response $J_{\textrm{em}}$, where $J_{\textrm{em}}$ describes the left hand side of Eq. (\ref{eq:holographic_design_reduced}), as $\alpha_0$ is varied from $\alpha_{0,\textrm{opt}}/2$ to $\alpha_{0,\textrm{opt}}$. In this case, the accessible phase increases from 180 degrees for the case of $\alpha_{0,\textrm{opt}}/2$, represented by a circle in the lower half plane, to a full 360 degrees at $\alpha_{0,\textrm{opt}}$. 

\begin{figure}[ht!]
\centering
\includegraphics[width=.7\columnwidth]{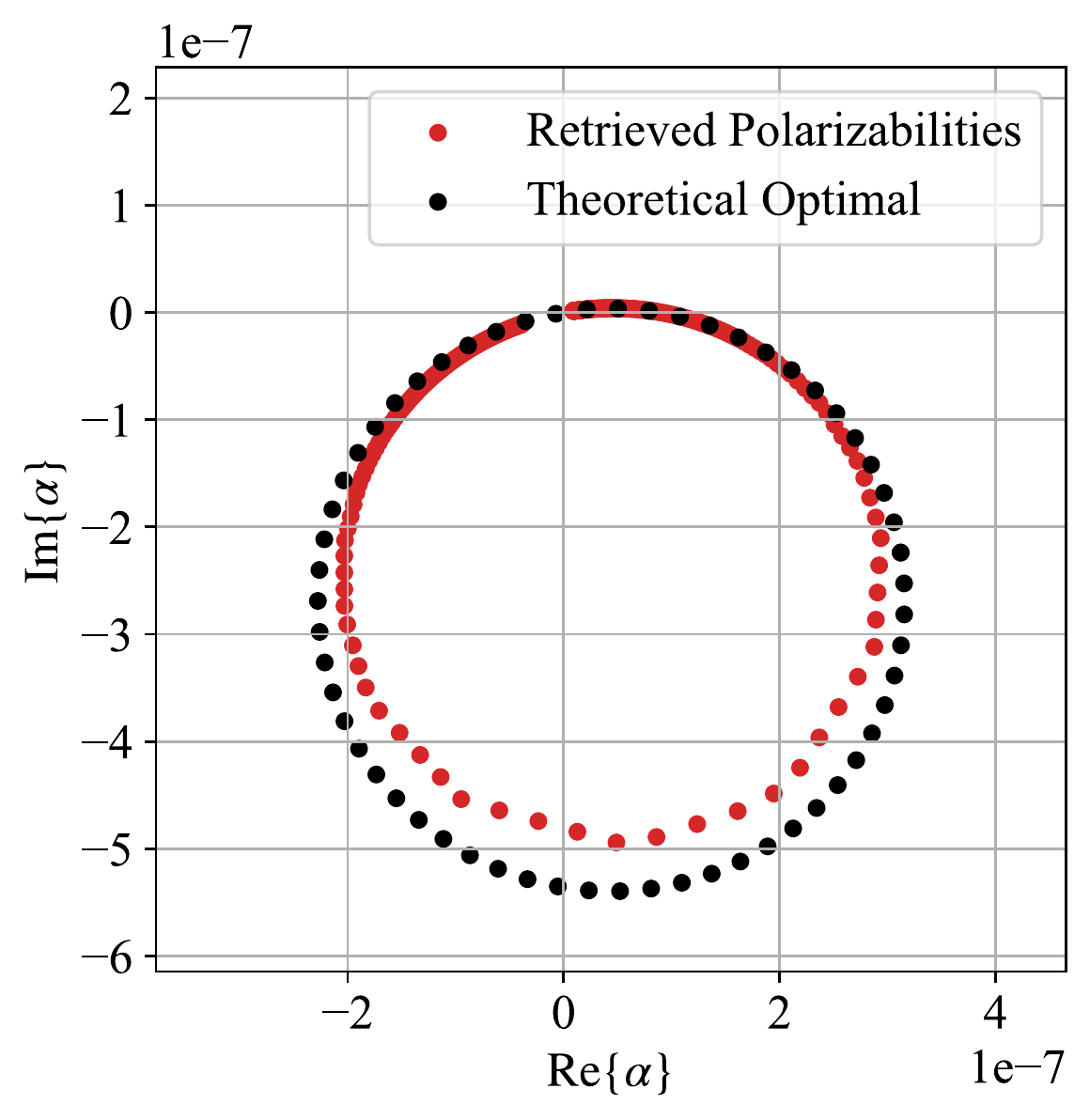}
\caption{Retrieved grayscale polarizabilities from Figs. \ref{fig:polarizability}(a) and (b) plotted in the complex plane (red), compared with the complex polarizability prescribed for optimal phase coverage by Eq. (\ref{eq:optimal_polarizability}) divided by $2(1-\Gamma)$ to isolate the contribution to the ideal magnitude from a single slot dipole.}
\label{fig:ideal_comparison}
\end{figure}

For the examples studied in the results that follow, $k_0$ and $\Lambda$ correspond to a frequency of 10 GHz and a lattice spacing of $\lambda/2$, and result in a polarizability magnitude of $\alpha_{0,\textrm{opt}}=2.14\times 10^{-6} \textrm{m}^3$ for optimal phase coverage. Using the substrate reflection coefficient $\Gamma$ and the approximation of an infinitesimal patch, the approximate effective polarizability of a single patch is comprised of contributions from two slot dipoles and two image dipoles, and so is found to be approximately $2(1-\Gamma)$ times the polarizability of a single slot dipole. Then a more accurate comparison for concluding optimality in the case of a ground plane-backed patch is $|\alpha_{0,\textrm{opt}}/(2(1-\Gamma))| = 5.43\times 10^{-7}\textrm{m}^3$, which approaches the peak effective polarizability magnitude for the grayscale values shown in Fig. \ref{fig:polarizability}(a) given by $4.97\times10^{-7}\textrm{m}^3$. The retrieved grayscale polarizabilities from Figs. \ref{fig:polarizability}(a) are plotted in the complex plane in Fig. \ref{fig:ideal_comparison} parameterized by patch length, alongside the optimal complex polarizability corresponding to a single dipole, $\frac{1}{2(1-\Gamma)}\frac{2\Lambda^2}{k_0}\left(\frac{-j+e^{-j\psi}}{2}\right)$, indicating the potential for successful beam formation using the retrieved polarizability values. 

\section{Results}
\label{sec:results}

\begin{figure}[ht!]
\centering
\includegraphics[width=\columnwidth]{"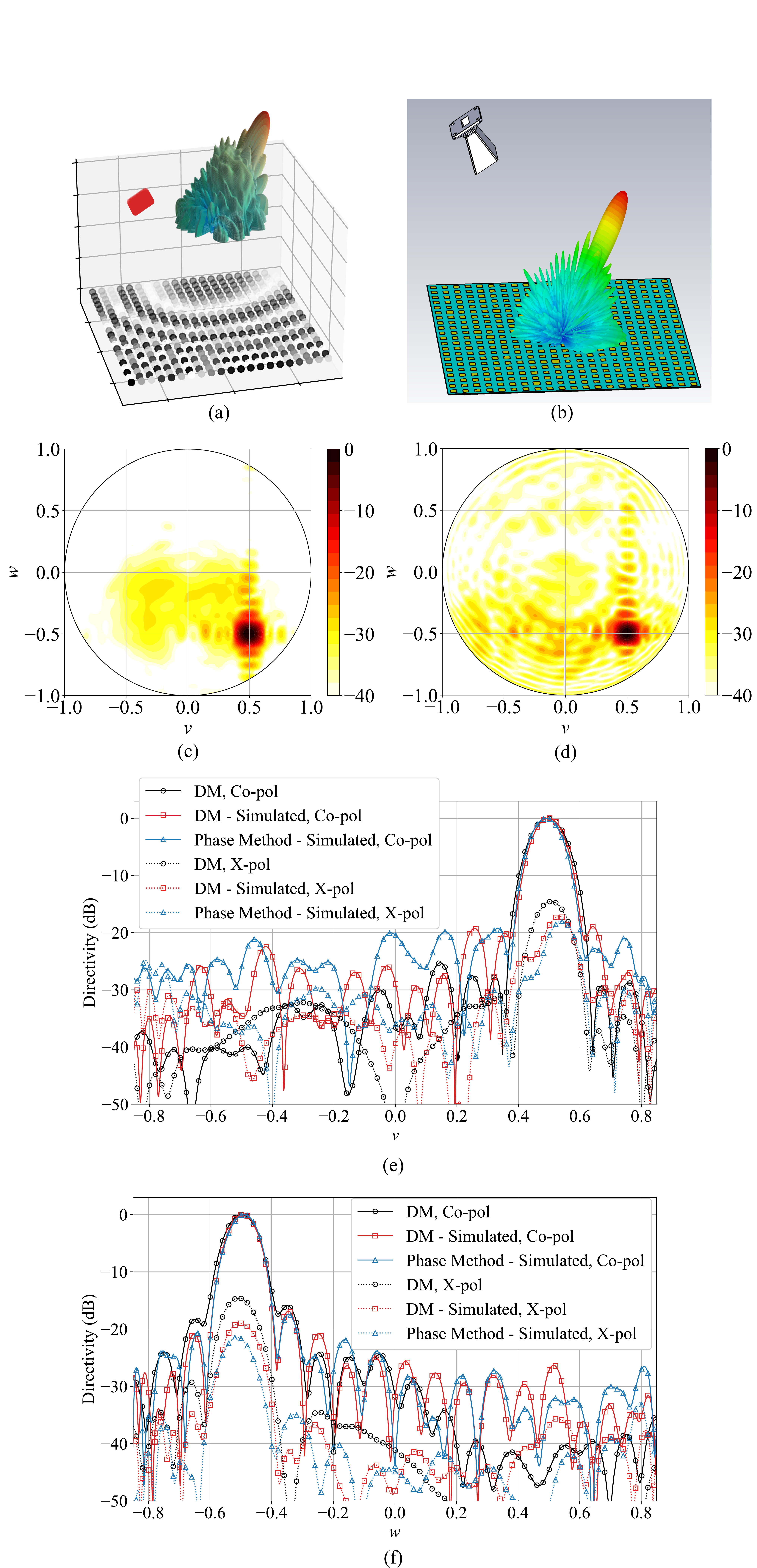"}
\caption{Comparison of far-field radiation patterns predicted by full-wave simulation software CST versus the proposed dipole model (DM) for a variable patch size reflectarray. (a) The 30 cm $\times$ 30 cm reflectarray geometry used for the grayscale design, shaded according to the real part of its grayscale polarizability value at each location. Also plotted are the positions of the horn aperture samples and the three-dimensional far-field beam. (b) The reflectarray model used for full-wave simulation, with patch dimensions chosen according to a minimum-length metric to steer a beam to $(v_0, w_0)=(0.5, -0.5)$. Contour plot of the far field pattern computed (c) directly by the DM, and (d) in full-wave simulation. (e)-(f) Beam cross sections along the $v$ and $w$ directions for the co-polarization (solid) and cross-polarization (dashed) components, including those computed directly by the DM, designed using the DM and computed in full-wave simulation (DM - Simulated), and designed using the ray tracing phase model and computed in full-wave simulation (Phase Method - Simulated).}
\label{fig:Grayscale-PM-CST}
\end{figure}

To demonstrate application of our polarizability model in the design of reflectarrays, we first consider the case of varying patch lengths with polarizabilities given in Figs. \ref{fig:polarizability}(a). The reflectarray in Fig. \ref{fig:Grayscale-PM-CST} covers a 30 cm $\times$ 30 cm area, with patches spaced at 1.5 cm, or half the free-space wavelength. A pyramidal horn antenna positioned (20 cm, -10 cm, 0) relative to the reflectarray center and rotated by $30^{\circ}$ about the $z$ axis illuminates the antenna with its magnetic field polarized in the $\hat{z}$ direction. The horn position and orientation were selected so that a majority of the horn's main beam resided within the aperture extent, and conforms to practical values that lead to satisfactory aperture efficiency \cite{pozar1997design, yu2010aperture}. The reflection coefficient $\Gamma$ was calculated according to 0.762 mm-thick Rogers 4350B substrate, and the polarizability distribution was selected in order to steer a beam to $(v_0, w_0)=(0.5, -0.5)$, where $v$ and $w$ are direction cosines defined with respect to the $y$ and $z$ axes (see Fig. \ref{fig:coordinates}). 

The design procedure consists of first specifying the patch array positions. Each patch consists of two polarizable slots oriented in the $\hat{z}$ direction. Although we model each patch with two magnetic dipoles, our assumption of geometrically symmetric patches requires that both of these dipoles be identical. Therefore, in order to impose this equality constraint during the design stage, we approximate the patch as a single dipole positioned at the center of the patch. We then apply Eq. (\ref{eq:polarizability_design}), here using a value of $H_{0z}=1$, and $H_{tz}$ numerically calculated at all array positions according to the illumination mode, here a pyramidal horn. 

\begin{figure}[ht]
\centering
\includegraphics[width=\columnwidth]{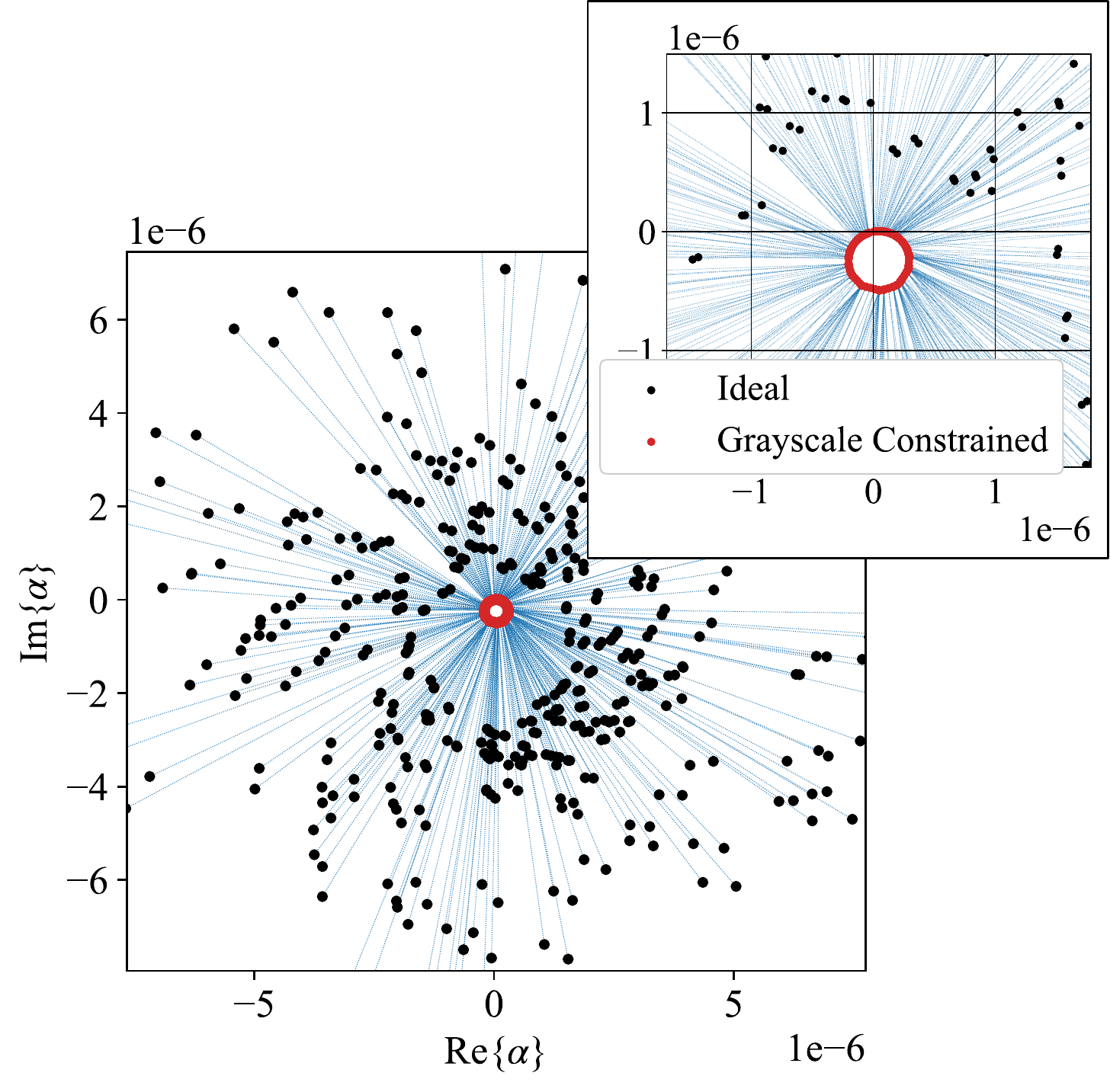}
\caption{Grayscale polarizability mappings in the complex plane for pyramidal horn illumination with the horn 20 cm from the reflectarray surface. Blue dots indicate the desired, unconstrained polarizability values returned by Eq. (\ref{eq:polarizability_design}), while the red dots represent the available polarizabilities retrieved for different patch sizes (Fig. \ref{fig:polarizability}(a)). The extent of the unconstrained values in the complex plane is adjusted by the scaling constant $a$.}
\label{fig:grayscale-polarizability-mapping}
\end{figure}

Equation (\ref{eq:polarizability_design}) yields unconstrained polarizability values at the patch locations that must then be constrained to the realizable polarizability values. For a grayscale demonstration, we take the values shown in Fig. \ref{fig:polarizability}(a) as our achievable design values, and map the unconstrained polarizabilities to a corresponding patch size according to a criterion of minimum length in the complex plane \cite{bowen2022optimizing}:
\begin{equation}
    \alpha_i = \textrm{arg}\min_{\alpha_c}\lVert \alpha_c - \alpha_{0,i} \rVert
\end{equation}
where $\lVert \cdot \rVert$ denotes the $l_2$-norm in the complex plane between the ideal polarizability prescribed for position $i$, $\alpha_{0,i}$, and the available patch polarizabilities $\alpha_c$. This constraint method is illustrated through a complex-plane representation in Fig. \ref{fig:grayscale-polarizability-mapping}, where the black dots correspond to the ideal polarizability values returned by Eq. (\ref{eq:polarizability_design}), and the red dots indicate the achievable polarizabilities of the variable-length patches from Fig. \ref{fig:polarizability}(a). The extent of the ideal polarizabilities in the complex plane has been modified by adjusting the scaling constant $a$ to heuristically realize improved design performance. The inset reveals the constrained polarizability states and resulting mapping in more detail. Note that the ideal polarizability values must compensate for the varying magnitude of the feed horn, forming a seemingly irregular pattern in the complex plane. Were we to ignore this magnitude variation, the ideal polarizabilities would instead lie on a circle in the complex plane.

Before making this mapping, we note that the polarizability value recovered using Eq. (\ref{eq:ff_extraction}) corresponds to that of a single slot. In order to properly make the correspondence between the prescribed polarizabilities at the patch locations and the slot polarizabilities of Eq. (\ref{eq:polarizability_design}), we multiply the slot polarizability values by a factor of $2 (1-\Gamma)$. This accounts for the presence of the two slot dipoles by an array factor approximation in the limit of an infinitesimally small patch $2\textrm{cos}\left(\frac{kW}{2} \textrm{sin}\theta \textrm{sin}\phi\right)\rightarrow 2$, as well as the slot image dipoles by an additional factor of $(1 - \Gamma)$. Once the constrained design is achieved, we replace the approximate array factor by the true array factor $(1-\Gamma)2\textrm{cos}\left(\frac{kW}{2} \textrm{sin}\theta \textrm{sin}\phi\right)$ for beam prediction and analysis, again assuming a thin substrate. The magnetic surface currents can then be computed according to Eq. (\ref{eq:Jm}) using the illuminating magnetic field at each patch position, and the radiation pattern calculated from the combined magnetic and electric surface currents.

Figure \ref{fig:Grayscale-PM-CST}(c) illustrates the co- and cross-polarization far field patterns resulting from our grayscale reflectarray design according to radiation integrals numerically applied to the equivalent magnetic and electric surface currents of our dipole model (DM), compared to that computed by the full-wave electromagnetic field solver CST\texttrademark (d), revealing that the polarizability model successfully predicts the main beam location as well as qualitative side lobe behavior. Co- and cross-polarizations are defined according to the third Ludwig definition \cite{ludwig1973definition, bhattacharyya2006phased}. The illumination in the full-wave simulation is provided by a WR-90 pyramidal horn excited by a rectangular waveguide port and positioned 20 cm from the reflectarray surface. For comparison, we also include simulated beam results obtained from a reflectarray designed using the conventional phase method of Eq. (\ref{eq:phase_mapping}), which resulted in a grayscale patch distribution noticeably distinct from, but qualitatively similar to, that obtained using the DM. The directivity computed using the polarizability model is 28.2 dBi, compared with 28.7 dBi as predicted by full-wave simulation of the DM design, and 27.7 dBi for the full-wave simulated phase method design. Meanwhile, The polarizability model reveals a 3-dB beamwidth of $8.7^{\circ}$, slightly larger than the beamwidths of $7.7^{\circ}$ and $7.5^{\circ}$ computed by full-wave simulation for the DM and phase models, respectively.

\begin{figure}[ht]
\centering
\includegraphics[width=\columnwidth]{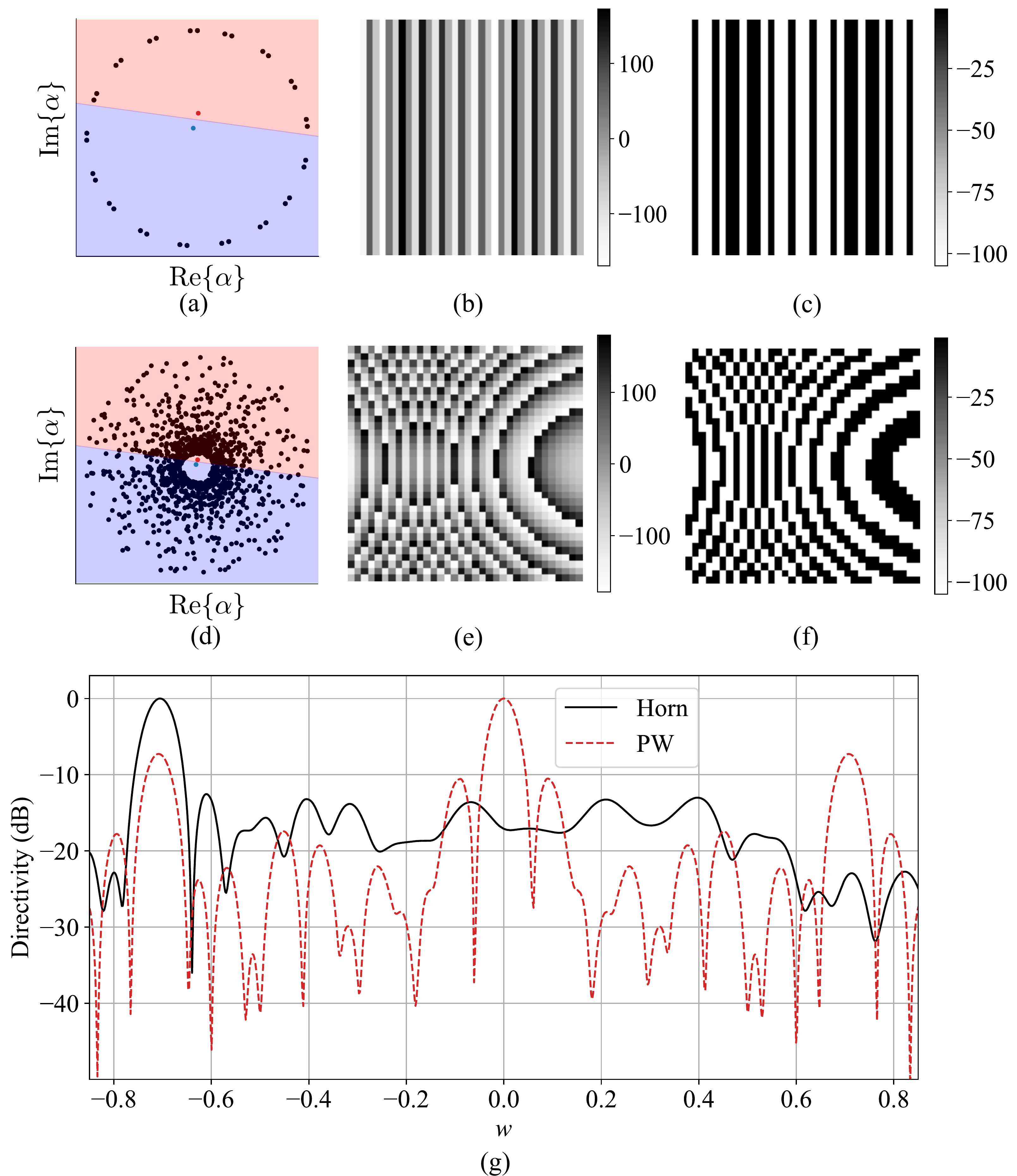}
\caption{Polarizability mapping and far field predictions for a 50 cm $\times$ 50 cm reflectarray under plane wave (PW) and horn illumination, steered to $(v_0, w_0)=(0, -0.7)$. (a) Complex-plane polarizability mapping illustrating the procedure of constraining the ideal polarizabilities (black) resulting from Eq. (\ref{eq:polarizability_design}) to a binary pair of available polarizabilities illustrated in Fig. \ref{fig:polarizability}(b) for plane wave illumination. Polarizability values in the half-plane shaded blue get mapped to the $on$ state (blue dot), while those in the red-shaded half-plane are mapped to the $off$ state (red dot). (b) The real part of the unconstrained polarizability distribution over the aperture, and (c) corresponding binary-constrained polarizability distribution. The corresponding polarizability mapping and spatial distributions for pyramidal horn illumination are shown in (d)-(f), where (e) depicts the real part of the unconstrained polarizabilities and (f) the binary-constrained polarizabilities. (g) compares the $w$ section between the binary-constrained radiation patterns under plane wave and horn illumination, where quantization lobes are particularly evident in the case of plane wave illumination.}
\label{fig:PW-horn}
\end{figure}

Figure \ref{fig:Grayscale-PM-CST} highlights the ability of the polarizability model to design and model patch reflectarrays with reduced computational complexity. In this case, the polarizability model accommodates grayscale patch design while incorporating both the amplitude and phase of the illuminating fields, as well as the amplitude and phase response of the patches themselves. These effects can be crucial in predicting performance under non-trivial illumination strategies. To demonstrate the effects of different illumination on beam performance and the ability of our polarizability model to capture these effects while further illustrating the polarizability selection procedure, Fig. \ref{fig:PW-horn} compares DM-predicted performance of a binary reflectarray under plane wave versus pyramidal horn illumination. The design utilizes a 50 cm $\times$ 50 cm apeture operating at 10 GHz. The horn antenna in this case is positioned 30 cm from the reflectarray and centered over the reflectarray. The horn aperture is oriented normal to the reflectarray surface, while the electric field polarization for both the plane wave and horn illumination points in the $\hat{y}$ direction.

Figure \ref{fig:PW-horn}(a) shows the complex-plane polarizability mapping between the unconstrained polarizability values obtained using Eq. (\ref{eq:polarizability_design}) and the binary values corresponding to the radiating (\textit{on}) and non-radiating (\textit{off}) patches for normal plane wave illumination. The corresponding constrained values are shown in Fig. \ref{fig:PW-horn}(c). For this demonstration, we take the values shown in Fig. \ref{fig:polarizability}(b) as our achievable design values, and map the unconstrained polarizabilities to these binary polarizabilities using a simple phase thresholding method. That is, for each patch position, we compare the phase of the unconstrained value obtained by Eq. (\ref{eq:polarizability_design}) to the phase of the radiating patch. If the phase difference is less than a chosen threshold, here taken as $90^{\circ}$, then a radiating patch is placed at that location. Otherwise, a non-radiating, shorted patch is used. In the complex-plane representation of the polarizabilities, this $90^{\circ}$ threshold maps the unconstrained values to the binary value residing in the same half-circle of the complex plane. The periodicity observed in the binary solution is known to result in strong quantization lobes \cite{theofanopoulos2020mitigating, yang2016study, smith1983comparison}, which are evident in the far-field radiation pattern illustrated in Fig. \ref{fig:PW-horn}(g). In contrast, the designed polarizability distribution obtained under pyramidal horn illumination similarly oriented with its magnetic field in the $\hat{z}$ direction is depicted in Fig. \ref{fig:PW-horn}(e) and its binary-constrained mapping in (f). The polarizability design corresponding to horn illumination exhibits noticeably reduced periodicity, which substantially improves the far field response (Fig. \ref{fig:PW-horn}(g)) by eliminating quantization lobes.

\begin{figure}[ht]
\centering
\includegraphics[width=.8\columnwidth]{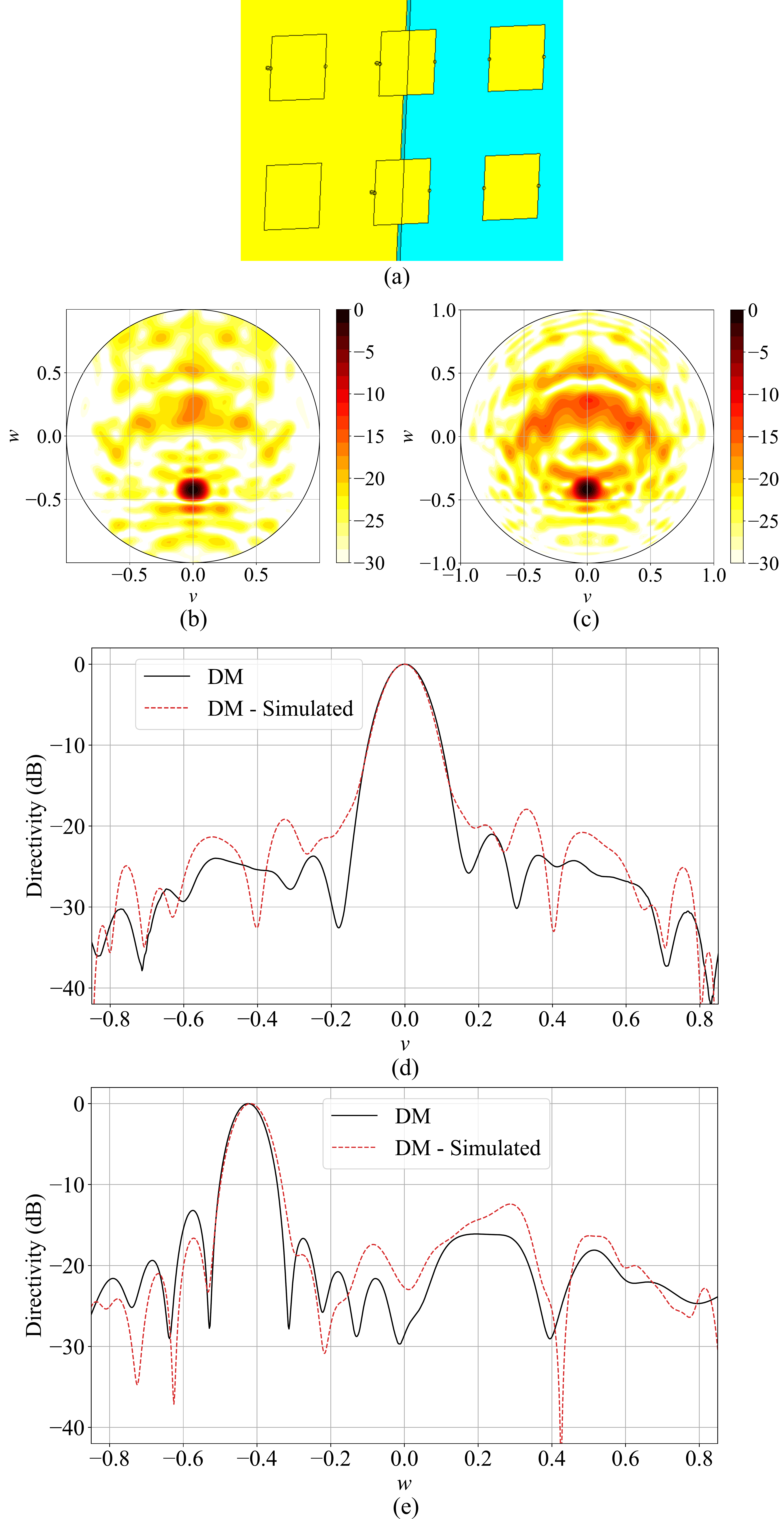}
\caption{Comparison of far-field radiation patterns predicted by full-wave simulation versus the DM for a binary-tuned reflectarray. (a) A close up view of the resonating and non-resonating patches with the dielectric partially removed. The vias used to make the patches non-resonating can be seen in four of the five patches shown. Far-field radiation pattern contour plots for the designed reflectarray computed through (b) the polarizability model and (c) full-wave simulation. Cross sections of the computed far-field patterns along the (d) $v$ and (e) $w$ directions at the peak location.}
\label{fig:PM-CST}
\end{figure}

In Fig. \ref{fig:PM-CST} we compare the beam pattern predicted by our polarizability model to that computed in the full-wave electromagnetic simulation software CST for a binary reflectarray. The designed reflectarray consists of the patches corresponding to the polarizability values reported in Fig. \ref{fig:polarizability}(b) and employed in the previous result of Fig. \ref{fig:PW-horn}. Here, the horn is oriented at the center of the reflectarray a distance 20 cm from and directed normal to the reflectarray surface. The patches are again spaced at half of the free space wavelength to cover a 30 cm $\times$ 30 cm area, and the polarizability distribution selected by Eq. (\ref{eq:polarizability_design}) to steer a beam to $(v_0, w_0) = (0, -0.42)$. After constraining the ideal polarizabilities to binary states using the previously described phase thresholding method, the designed geometry was modelled and simulated in the frequency domain to evaluate the far field response. The resonating and shorted patch geometries can be seen in detail in Fig. \ref{fig:PM-CST}(a), which provides a closeup perspective of six of the modelled patches with the dielectric substrate partially removed. The far-field radiation pattern computed by our polarizability model is illustrated in the contour plot of Fig. \ref{fig:PM-CST}(b), which we may compare to that obtained from full-wave simulation of the design (c). Figures \ref{fig:PM-CST}(d)-(e) show cross sections over $v$ and $w$ at the beam peak direction, and reveals that the beam profile predicted using the polarizability model agrees well with full-wave simulation. The corresponding peak directivity and 3-dB beamwidth values were calculated as 26.5 dBi and $5.9^{\circ}$ for the pattern computed with our polarizability model, and 25.7 dBi and $6.1^{\circ}$ for the the simulated result, which agree well.

\section{Discussion}\label{sec:discussion}
The results of the previous section reveal qualitative agreement between the polarizable patch model and more rigorous full-wave strategies, yielding satisfactory main beam, directivity and beamwidth estimates. Whereas full-wave simulations of the full reflectarray structures required several hours of computation, the discrete dipole method achieved comparable results in seconds. Nevertheless, while sidelobe levels agree approximately, it is evident from Figs. \ref{fig:Grayscale-PM-CST} and \ref{fig:PM-CST} that the sidelobe profiles predicted by the DM do not generally adhere to those obtained through simulation. We believe that improvements in accuracy can be realized upon discarding the weakly scattering approximation and implementing a fully coupled dipolar model \cite{yoo2019analytic2}, including both magnetic, electric, and magnetoelectric interactions between the patch magnetic dipoles and ground plane electric currents \cite{bowen2012using, mulholland1994light}. While the presented studies used sufficiently separated patches to minimize magnetic dipole coupling, the interaction between magnetic dipoles and the ground plane was accounted for in this work simply using image theory, which can give inaccurate results due to edge effects of the finite ground plane. A dipolar representation of the ground plane can provide a self-consistent treatment with improved accuracy. Similarly, only the fields over the horn aperture were accounted for in the present work, and scattering from the horn was not considered. Incorporating these effects into a dipolar model will be the subject of future research. 

While we restricted our study to the subject of polarizable, rectangular microstrip patches, the proposed methods can equally well be applied to patches and metamaterial elements of arbitrary geometry, as long as they are sufficiently subwavelength to permit the dipole approximation. The coupled dipole model (Eq. (\ref{eq:DDA})) enables efficient and accurate computational modeling of large-scale reflectarrays with closely spaced, subwavelength elements that offer advantages in bandwidth improvement \cite{pozar2007wideband, nayeri2011bandwidth} and improved beam performance under quantization constraints \cite{yang2016study}. Regarding bandwidth, we have restricted our study to a single operating frequency to emphasize the polarizability extraction and antenna design procedures. Nevertheless, we point out that reflectarray bandwidth limitations should not be modified by operating under the polarizability framework, as bandwidth is fundamentally determined by physical resonance characteristics of which the polarizability is just one description. Past work in our lab confirms that the dipolar model of reflectarrays results in behavior corresponding to that found in the literature \cite{huang1995bandwidth, pozar2003bandwidth}.

Finally, although this work has investigated application of the discrete dipole method to reflectarrays for beam steering, a straightforward modification of Eq. (\ref{eq:holographic_design}) allows accommodation of a much wider range of applications. Sufficient polarimetric control over the illumination and patch scattering behavior (contained in its polarizability tensor \cite{liu2016polarizability}) enables the synthesis of arbitrary aperture fields for applications such as contoured beams \cite{encinar2004three}, multi-beam antennas \cite{nayeri2011design}, and dynamic wavefront control \cite{liang2015reconfigurable}.

\section{Conclusion}\label{sec:conclusion}
In this work we have formulated a reflectarray design approach that models patches as pairs of polarizable magnetic dipoles, following a standard cavity model of the ground plane-backed patch antenna. The polarizabilities describing a patch can be recovered through numerical application of standard surface equivalence principles, or through numerical or experimental far-field measurements. Using these recovered polarizabilities, the patch reflectarray can be specified using a holographic inverse method. We have applied this method to the design of a variable patch size, grayscale state reflectarray, as well as that of a binary state reflectarray, and shown that the method results in accurate beam pattern predictions that agree reasonably well with full-wave simulation.

\section*{Acknowledgment}
This work was supported by the National Science Foundation award ECCS-2030068, NASA grant 80NSSC21M0197, and the NASA Game Changing Development (GCD) ARMADAS Project.

\ifCLASSOPTIONcaptionsoff
  \newpage
\fi

\bibliographystyle{unsrt}
\bibliography{reflectarray}




\end{document}